\documentclass[twocolumn,twocolappendix,iop,apj]{emulateapj}
\usepackage{times}
\usepackage{comment}
\usepackage{hyperref}
\usepackage[T1]{fontenc}
\usepackage{amsmath,wasysym}
\usepackage{natbib}
\usepackage{xcolor}
\usepackage[normalem]{ulem}
\newcommand\bu{\bgroup\markoverwith{\color{blue}{\rule[-0.5ex]{2.5pt}{0.6pt}}}\ULon}
\newcommand\rs{\bgroup\markoverwith{\color{red}{\rule[0.5ex]{2pt}{0.6pt}}}\ULon}

\newcommand{\fermilat}{{\em Fermi}-LAT}

\newcommand{\gray}{$\gamma$-ray}

\newcommand{\GP}{{\it GALPROP}}

\begin{document}

\title{Cosmic-Ray Propagation in Light of Recent Observation of Geminga}


 \author{Gu{\dh}laugur J{\'o}hannesson}
 \affiliation{Science Institute, University of Iceland, IS-107 Reykjavik, Iceland}
 \affiliation{Nordita, KTH Royal Institute of Technology and Stockholm University, Roslagstullsbacken 23, SE-106 91 Stockholm, Sweden}
\author{Troy A. Porter} \author{Igor V. Moskalenko}
\affiliation{W. W. Hansen Experimental Physics Laboratory and Kavli Institute for Particle Astrophysics and Cosmology, \\ Stanford University, Stanford, CA 94305, USA}


\begin{abstract}
The High Altitude Water Cherenkov (HAWC) telescope recently observed extended emission around the Geminga and PSR~B0656+14 pulsar wind nebulae (PWNe).  These observations have been used to estimate cosmic-ray (CR) diffusion coefficients near the PWNe that appear to be more than two orders of magnitude smaller than that typically derived for the interstellar medium from the measured abundances of secondary species in CRs. Two-zone diffusion models have  been proposed as a solution to this discrepancy, where the slower diffusion zone (SDZ) is confined to a small region around the PWN. Such models are shown to successfully reproduce the HAWC observations of the Geminga PWN while retaining consistency with other CR data.  It is found that the size of the SDZ influences the predicted positron flux and the spectral shape of the extended \gray{} emission at lower energies that can be observed with the {\it Fermi} Large Area Telescope (\fermilat). If the two observed PWNe are not unique, then it is likely that there are similar pockets of slow diffusion around many CR sources elsewhere in the Milky Way. The consequences of such picture for Galactic CR propagation is explored.

\end{abstract}

\section{Introduction}

The sources of majority of cosmic rays (CRs) are believed to be supernovae
(SNe) and supernova remnants (SNRs), which are capable of accelerating
particles to multi-PeV energies. As CRs propagate through the interstellar
medium (ISM) they scatter off the magnetic turbulences in a process that on large
scales can be described with a diffusive transport equation
\citep[][]{1964ocr..book.....G,1990acr..book.....B}. The spectrum of
turbulence controls the value of the diffusion coefficient and its energy
dependence because the CR particles most efficiently scatter on turbulence that is
comparable in size to their gyroradii. The turbulence is
assumed to initially form at large scales and then cascade down to smaller
scales thus affecting particles at all rigidities. The power-law shape of the
energy spectrum of turbulence
\citep{1941DoSSR..30..301K,1964SvA.....7..566I,1965PhFl....8.1385K} translates
into a power-law rigidity dependence of the diffusion coefficient with the
index taking a value between 0.3 and 0.6. Observations of the ratios of stable secondary to
primary species in CRs and the abundances of radioactive secondaries are
usually employed to determine both the power-law index and normalization of
the diffusion coefficient averaged over a significant volume of the
interstellar space surrounding the Solar system
\citep[e.g.,][]{1998ApJ...509..212S,2007ARNPS..57..285S,JohannessonEtAl:2016},
typically several kpc in radius. Recent estimates of the power-law index are
clustered around 0.35 with a normalization of about $4\times10^{28}$ cm$^2$
s$^{-1}$ at a rigidity of 4 GV \citep[e.g.,][]{BoschiniEtAl:2018b}.

Pulsars are rapidly spinning and strongly magnetized neutron stars that are at
the final stage of the stellar evolution. They are formed in SN explosions
and can often be found inside their associated SNR. 
Pulsars represent a class of CR sources that has not been considered as extensively as the more usual scenario of acceleration in SNR shocks, but the fact that they may produce CRs is well-known \citep{1981IAUS...94..175A,1987ICRC....2...92H,1989ApJ...342..807B}. However,
recent measurements of positrons in CRs 
\citep{AdrianiEtAl:2009, AckermannEtAl:2012, AguilarEtAl:2014}
in excess of predictions of propagation models
\citep{1982ApJ...254..391P,1998ApJ...493..694M}, made under an assumption of
their entirely secondary production in the ISM, elevated
pulsars to be one of the primary candidate sources responsible for this excess
\citep[e.g.,][]{2009JCAP...01..025H,2009PhRvL.103e1101Y,2009PhRvD..80f3005M}.
The magnetospheres of rapidly
rotating neutron stars are capable of producing electrons and positrons in
significant numbers and accelerating them to very high energies resulting in a
pulsar wind nebula (PWN) that is observable from radio to high-energy \gray{s} \citep{GaenslerSlane:2006}; an archetypical example of such a source is the Crab pulsar and its PWN. The accelerated particles eventually escape from the PWN into the ISM and some of them can reach the Solar system. Therefore, PWNe can
be natural candidates responsible for the puzzling excess in CR positrons
observed by several experiments.

Recent observations of the extended TeV emission around Geminga and PSR~B0656+14
PWN by the High Altitude Water Cherenkov (HAWC) telescope constrain the
diffusion coefficient in their vicinities to be about two orders of
magnitude smaller than the average value derived from observations of CRs
\citep{AbeysekaraEtAl:2017}. Application of such slow diffusion to the local
Milky Way results in a contradiction with other CR observations, in particular
observations of high-energy CR electrons and positrons.  Fast energy losses of
TeV particles through inverse Compton (IC) scattering and synchrotron emission
limits their lifetime to $\sim$100~ky \citep{1998ApJ...509..212S}.
If such slow diffusion is representative for the ISM
within about a few hundred pc of the Solar system, the sources of the TeV particles detected at Earth also need to be within a few 10s of pc.
\citet{ProfumoEtAl:2018} highlighted that such nearby sources have
not been identified and proposed a
two-zone diffusion model with the slow diffusion confined to a small region
around the PWN. Other authors have considered similar scenarios with varying
details \citep{TangPiran:2018,FangEtAl:2018,EvoliEtAl:2018}. Interestingly, a
similar suppression of the CR diffusion was observed in the Large Magellanic
Cloud around the 30 Doradus star-forming region, where an analysis of combined
\gray{} 
and radio observations 
yielded a diffusion coefficient, averaged over a region with radius
200--300 pc, an order of magnitude smaller than the typical value in the Milky
Way \citep{2012ApJ...750..126M}. Meanwhile, a strong suppression of the
diffusion coefficient around a SNR due to excitation of magnetic
turbulence by escaping CRs
was predicted some time before the HAWC observations were reported \citep[e.g.,][see also references therein]{2008AdSpR..42..486P,MalkovEtAl:2013,2016PhRvD..94h3003D}. A similar mechanism may also be at work in PWN.

In this paper the two-zone diffusion model is explored using the latest
version of the GALPROP\footnote{\url{http://galprop.stanford.edu} \label{url}}
propagation code \citep{PorterEtAl:2017,JohannessonEtAl:2018}.  The results indicate that such a model is a viable interpretation for the HAWC observations and confirm similar conclusions made by other authors. Predictions for lower-energy \gray{} emission that can be observed with the {\it Fermi} Large Area Telescope
(\fermilat) are made and the contribution of energetic positrons coming from Geminga to the
observed CR positron flux in different scenarios is studied. Effects of the
size of the slower diffusion zone (SDZ) and the properties of the accelerated electrons/positrons are taken into
account as is the effect of the proper motion of Geminga. Unless
both Geminga and PSR~B0656+14 are special cases there are expectations for more regions of
slower diffusion around other PWNe in the Milky Way.  The implications that
such inhomogeneity of the diffusion in the ISM can have on the CR distribution
throughout the Milky Way is also explored.

\section{A Model for Geminga}

\subsection{Physical Setup}

It is assumed that the Geminga pulsar is injecting accelerated electrons and
positrons into the ISM in equal numbers with a fraction $\eta$ of its
spin-down power converted to pairs.   After injection the particles
propagate via a diffusive process.  The pulsar
parameters used for this paper are identical to those from
\citet{AbeysekaraEtAl:2017} and the
energy distribution of the injected electrons/positrons is described with a
smoothly joined broken power-law
\begin{equation}
  \frac{dn}{dp} \propto E_k^{-\gamma_0}\left[ 1 + \left( \frac{E_k}{E_b}
  \right)^\frac{\gamma_1-\gamma_0}{s} \right]^{-s}.
  \label{eq:injectionSpectrum}
\end{equation}
Here $n$ is the number density of electrons/positrons, $p$ is the particle
momentum, $E_k$ is the particle kinetic energy, and $\gamma_1$ is a
power-law index at high energies.  The smoothness parameter $s=0.5$ is assumed
constant and so is the low-energy index $\gamma_0 = -1$ and the break energy
$E_b=10$~GeV. This low-energy break effectively truncates the injected particle spectrum that is not expected to extend unbroken to lower energies \citep[e.g.,][]{Amato:2014}.  The break is required to keep the value of $\eta$ below 1 for the largest values of $\gamma_0$ considered.  The truncation occurs at energies below that explored in this paper and has no effect on the results.
The injection spectrum is normalized so that the total power injected is given by the expression
\begin{equation}
  L(t) = \eta \dot{E}_0 \left( 1 + \frac{t}{\tau_0} \right)^{-2},
  \label{eq:PulsarPower}
\end{equation}
where $\dot{E}_0$ is the initial spin-down power of the pulsar, and $\tau_0 =
13$~kyr.  The initial spin-down power is calculated using the current
spin-down power of $\dot{E}=3.26\times10^{34}$ erg s$^{-1}$ assuming that the
pulsar age is $T_p = 340$~kyr.  The distance to Geminga has been determined to
be 250~pc \citep{FahertyEtAl:2007}. The spatial grid for the propagation
calculations with GALPROP is right handed with the Galactic centre (GC) at the origin, the Sun
at $(x,y,z)=(8.5, 0, 0)$~kpc, and the $z$-axis oriented toward the Galactic
north pole. This coordinate system places Geminga at $(8.7407, 0.0651, 0.0186)$~kpc at the current epoch.

Previous interpretations of the HAWC observations have assumed that the Geminga
pulsar is a stationary object
\citep{ProfumoEtAl:2018,TangPiran:2018,FangEtAl:2018,EvoliEtAl:2018}.  Only
\citet{TangPiran:2018} discussed the effect of the proper motion of Geminga,
but concluded that it has little effect on electrons generating the observed
TeV \gray{} emission because of their short cooling timescale.
\citet{FahertyEtAl:2007} measured the proper motion of Geminga to be 107.5 mas
yr$^{-1}$ in right ascension and 142.1 mas yr$^{-1}$ in declination.  Transforming this into
Galactic coordinates leads to $-80.0$ mas yr$^{-1}$ in longitude and 156.0 mas
yr$^{-1}$ in latitude.
Assuming a constant proper velocity that is currently
perpendicular to the line-of-sight, the corresponding velocity in the spatial
grid coordinate system is $(v_x, v_y, v_z) = (24.3, -89.9,
182.6)$~km s$^{-1}$ with Geminga originally located at $(8.7320, 0.0963, -0.0449)$~kpc.  
In this case, Geminga was born in a SN explosion of a star that
has travelled from the direction of the Orion OB1a association
\citep{SmithEtAl:1994}. While the TeV \gray{} emission is not significantly
affected by the proper motion, the large traveled distance and proximity of
Geminga to the Sun should produce a ``trailing'' tail of CRs whose \gray{} emissions may be observable
at lower energies with \fermilat.

For the two-zone diffusion model, the diffusion coefficient in a confined
region around the pulsar is assumed to be lower than that in the ISM due to
the increased turbulence of the magnetic field.
This region will hereafter be called a ``slower diffusion zone'' (SDZ). 
It is also assumed that the stronger turbulence does not change the power
spectrum and hence the rigidity dependence of the diffusion coefficient is the
same for the SDZ and the ISM. Let $r$ be the distance from the center of the
SDZ, then the spatial dependence of the diffusion coefficient is given by
\begin{equation}
  D\!=\!\beta \left( \frac{R}{R_0} \right)^\delta
  \begin{cases}
    \displaystyle D_z, & r < r_z, \\
    \displaystyle D_z \left[ \frac{D_0}{D_z} \right]^{\frac{r-r_z}{r_t-r_z}}, & r_z \le r \le r_t, \\
    D_0, & r > r_t. 
  \end{cases} 
  \label{eq:diffusion}
\end{equation}
Here $\beta$ is the particle velocity in units of the speed of light,
$D_0$ is the normalization of the diffusion coefficient in the general ISM,
$D_z$ is the normalization of the diffusion coefficient within the SDZ
with radius $r_z$, $R$ is the particle rigidity, and $R_0=4$ GV is the normalization (reference) rigidity. In the transitional layer between $r_z$ and $r_t$, the
normalization of the diffusion coefficient increases exponentially with $r$
from $D_z$ to the interstellar value $D_0$. Depending on the exact origin of
the SDZ, its radius can be time dependent as can its location. To be
consistent with the HAWC observations, both $r_z$ and $r_t$ have to be of the
order of a few tens of pc at the current time.

The effect of the SDZ around the PWN on the spectra and distribution of
injected electrons/positrons and their emission may depend on its origin. 
Two general categories distinguished by the origin of the SDZ are
considered. For the first category the SDZ is due to events external to the PWN itself  (external SDZ) that may be related to the
progenitor SNR, or surrounding environment.  It is assumed that the
particle propagation time in the zone is much longer than the time necessary
to generate such a zone itself (``instant'' generation) and its location is
fixed.  For the second category, the SDZ is assumed to be associated with and
generated by the PWN itself (PWN SDZ).
Therefore, the SDZ is moving with the Geminga pulsar and its size increases
proportionally to the square root of time. This is in qualitative
agreement with the evolution of a PWN as determined from simulations
\citep[e.g.][]{vanderSwaluw:2003}.

For the PWN SDZ, the expectation is that there may be a link
between the pulsar that is generating the magnetic turbulence and the surrounding
interstellar plasma. In this case, the pulsar would transfer a momentum to the medium within the SDZ.
To estimate the evolution of the velocity of Geminga a simple model is put forward. It is assumed that
the radius of the SDZ region $r_z$ around Geminga is increasing in a diffusive manner so that $r_z(t) = \mu \sqrt{t}$, where $\mu$ is a constant. It is further assumed that a fraction $f$ of the ISM within
$r_z$ is swept by the PWN and sped up to the velocity $v$ of the pulsar. Using conservation of momentum $dp = -v dM$ gives
$v=v_0 (M_0/M)^2$, where $v_0$ and $M_0\approx M_\odot$ are the initial velocity and mass of the system, respectively.
Substitution of $r_z(t)$ and $v(M)$ into the mass conservation formula $dM = f \pi \rho r_z^2 v dt$, yields the time evolution of the velocity of Geminga:
\begin{equation}
  v = \frac{v_0}{\left(3 A_d v_0 t^2 /2 + 1\right)^{2/3}}.
  \label{eq:vDiffusion}
\end{equation}
Here $A_d = f \pi \rho \mu^2 / M_0$ is the drag coefficient and $\rho \approx
0.03$~$M_\odot$~pc$^{-3}$ is the mass density of the ISM.
Eq.~(\ref{eq:vDiffusion}) can be integrated to get the full traveled distance:
\begin{equation}
 \lambda(t) = v_0 t\ _2F_1\left( \frac{1}{2}, \frac{2}{3}; \frac{3}{2}; -\frac{3 A_d v_0 t^2}{2} \right),
  \label{eq:lDiffusion}
\end{equation}
where $_2F_1$ is the hypergeometric function. Further assuming that $r_z(T_p)
\approx 30$~pc and that $f\approx 10^{-3}$, the distance traveled by Geminga
is $\lambda(T_p) \approx 330$~pc and its initial velocity is $v_0 \approx
4300$~km s$^{-1}$. This places the birth of Geminga at the location of Orion
OB1a, which is at a distance of 330~pc from its current location. 
This is a very intriguing possibility, but in this case $v_0$ is much higher
than the observed velocities of any other neutron stars that reach only up to
1000~km s$^{-1}$ \citep[e.g.,][]{HobbsEtAl:2005}.  Assuming $f\approx 10^{-4}$
results in a more reasonable speed of $v_0 \approx 400$~km s$^{-1}$ and a
total distance traveled of $\lambda(T_p) \approx 100$~pc.  A moderate slow
down agrees with observations of older pulsars having, on average, smaller velocities than young pulsars \citep{HobbsEtAl:2005}.

\subsection{Calculation Setup}

The calculations are made using the latest release of the GALPROP
code\textsuperscript{\ref{url}} \citep{PorterEtAl:2017,
JohannessonEtAl:2018}.  The GALPROP code solves the diffusion-advection
equation in three spatial dimensions allowing for diffusive re-acceleration in
the ISM -- see the website and GALPROP team papers for full details.  Of major
relevance for this paper, the code fully accounts for energy losses due to synchrotron
radiation and inverse Compton (IC) scattering. The resulting synchrotron and
IC emission is calculated for an observer located at the position of the Solar
system.  The IC calculations \citep{2000ApJ...528..357M} take into account the
anisotropy of the interstellar radiation field (ISRF). The current
calculations employ the magnetic field model of \citet{SunEtAl:2008} described
by their Eq.~(7) and the R12 model for the ISRF developed by
\citet{PorterEtAl:2017}. While the turbulent magnetic field
is expected to be larger in the SDZ, the magnetic field model is not updated
to account for this. The expected increase in the synchrotron cooling rate and
corresponding electromagnetic emissions will not significantly affect the results presented below.

The code has also been enhanced to use
non-equidistant grids to allow for increased resolution in particular areas of
the Milky Way, in this case around Geminga.
This improvement to \GP\ is inspired by the the Pencil Code\footnote{See Section 5.4 of
  \url{http://pencil-code.nordita.org/doc/manual.pdf}}
\citep{BrandenburgDobler:2002}, where the usage of analytic functions can have advantages in terms of speed and memory usage compared to purely numerical implementations for non-uniform grid spacing.
The current run uses the grid function
\begin{equation}
  z(\zeta) = \frac{\epsilon}{a}\tan\left[ a\left( \zeta - \zeta_0 \right)  \right] + z_0
  \label{eq:gridFunction}
\end{equation}
for all spatial coordinates $\zeta = x, y, z$, where $\epsilon, a, \zeta_0, z_0$
are parameters. This function maps from the linear grid $\zeta$ to the
non-linear grid in each of the spatial coordinates $x, y, z$. The equations are solved on the $\zeta$ grid after the
differential equations have been updated to account for the change in first
and second derivatives.

The parameters of the grid function are chosen such that the resolution is 2
pc at a central location that is defined below for each calculation, but goes
up to $0.1$\,kpc at a distance of $700$\,pc from the grid center.  This setup
provides about 0.2 degree resolution on the sky for objects
located at a distance of 250~pc along the line of sight towards the center of the
grid.  To minimize artificial asymmetry, the grid
has the current location of Geminga close to the center of
a pixel and close enough to the center of the non-equidistant grid so that
there is little distortion due to the variable pixel size.  For the chosen
parameters the grid size is approximately uniform within a box having a width of
$\sim60$~pc, which is not enough to enclose the entire evolution of the
Geminga PWN.  The resolution and size of the grid is bounded by computation costs; the selected parameters are a result of a compromise
between accuracy, memory requirements, and computation speed.  This
necessarily means that the start and (or) the end of the evolution of Geminga are (is) not fully resolved.  Because the diffusion process smooths the distribution of particles as time evolves, the grid is chosen such that the current location of Geminga, and hence the TeV \gray{} emission, is
always well resolved, but sometimes this may come with a small decrease of spatial resolution at its birth place.

The calculations are performed in a square box with a width of 8~kpc.
This is wide enough so that the CR propagation calculations in the Solar neighborhood are not affected by the boundary conditions. The non-equidistant grid allows the boundaries to be extended this far without imposing large computational
costs.  A fixed timestep of 50 years is
used for the calculations.  This is small enough to capture the propagation and
energy losses near the upper boundary of the energy grid, which is 1~PeV. The
upper energy boundary is chosen to be well above that for the particles
producing the HAWC data.  The lower energy
boundary is set at 100 MeV, much lower than the cutoff in the injection
spectrum.  The energy grid is logarithmic with 16 bins per decade.

The values of $D_0 = 4.5 \times 10^{28}$ cm$^2$ s$^{-1}$ and $\delta=0.35$ are
chosen to match the latest AMS-02 data on secondary CR species (see Section~\ref{sec:MW}). The SDZ diffusion coefficient of $D_z = 1.3 \times 10^{26}$ cm$^2$ s$^{-1}$ at $R_0=4$ GV corresponds to the value derived from the HAWC observations at higher energies \citet{AbeysekaraEtAl:2017}.  No attempt is made to independently fit the latter to the HAWC observations because each propagation calculation is
computationally expensive, taking $\sim$ 24
hours to complete on a modern 40 core CPU. The calculations also include
diffusive re-acceleration with an Alfv{\'e}n speed $v_A = 17$~km s$^{-1}$, as determined from the fit to the secondary-to-primary data.  
To save CPU time the calculations are made using electrons only,
because the energy losses and propagation of positrons and electrons are
identical.  When comparing to the measured positron flux at Earth, the
calculated particle flux is simply divided by two.  The IC emission is
calculated on a HEALPix \citep{GorskiEtAl:2005} grid having an order of 9
giving a resolution of about 0.1 degrees.  The IC emission is evaluated on a
logarithmic grid in energy from 10~GeV to 40~TeV with 32 energy planes.

\subsection{Results}

\begin{figure}[tb]
  \centering
  \includegraphics[width=0.48\textwidth]{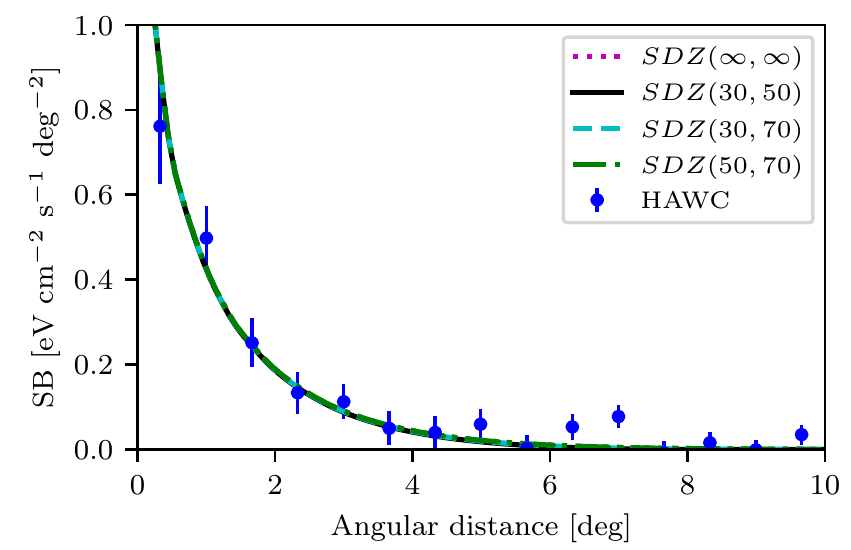}\\
  \includegraphics[width=0.48\textwidth]{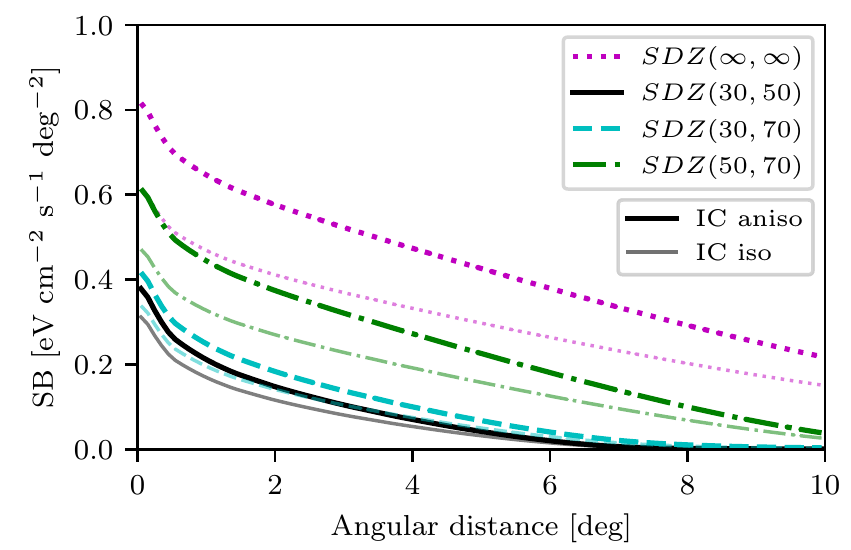}
  \caption{Surface brightness of the modelled IC emission shown as a function
    of angular distance from the current location of Geminga. The models shown
    here assume Geminga is stationary, $\gamma_1=2.0$, and the SDZ is
    stationary and centered at Geminga.  The top panel shows the energy range
    8~TeV to 40~TeV compared to the observations of HAWC
    \citep{AbeysekaraEtAl:2017}. The bottom panel shows the model predictions
  for the energy range 10~GeV to 30~GeV. The different curves represent
different assumptions about the size of the SDZ, encoded as SDZ($r_z$, $r_t$)
in the legend.}
  \label{fig:StationarySB}
\end{figure}

For a comparison with results of previous studies the first set of calculations is performed
assuming that there is no proper motion of Geminga.  The index of the injection spectrum is
fixed to $\gamma_1 = 2.0$ and the efficiency parameter is $\eta=0.26$. The SDZ
is centered at the current location of Geminga $(l_G, b_G) = (195.\!\!^\circ14, 4.\!\!^\circ26)$ and is static in size. The spatial grid is also centered
at the same location. Several combinations of $(r_z,r_t)$ are used,
covering a range 
from 30~pc to 50~pc for $r_z$ and from 50~pc to 70~pc for $r_t$. 
Calculations are also made using a model with $(r_z,r_t) =
(\infty, \infty)$ for a comparison with the results from
\citet{AbeysekaraEtAl:2017}. Figure~\ref{fig:StationarySB} (top panel) shows the angular
profile of the surface brightness of the modeled IC emission evaluated over the
energy range from 8~TeV to 40~TeV and compared with the data from HAWC.  The
resulting profile is clearly independent of the size of the SDZ,
because the cooling time limits the diffusion length at the corresponding CR
electron energies ($\gtrsim 100$~TeV) to be $\lesssim 10$~pc.  The same cannot be said for the IC emission evaluated
for the energy range from 10~GeV to 30~GeV, which is also shown in
Figure~\ref{fig:StationarySB} (bottom panel).  Because of the longer cooling timescales the
emission profile is highly sensitive to the size of the SDZ, via both $r_z$
and $r_t$. A smaller SDZ
size leads to a correspondingly lower number density of CR electrons/positrons within the zone, which produces a flatter profile and fainter emission in the GeV energy
range.

\begin{figure}[tb]
  \centering
  \includegraphics[width=0.48\textwidth]{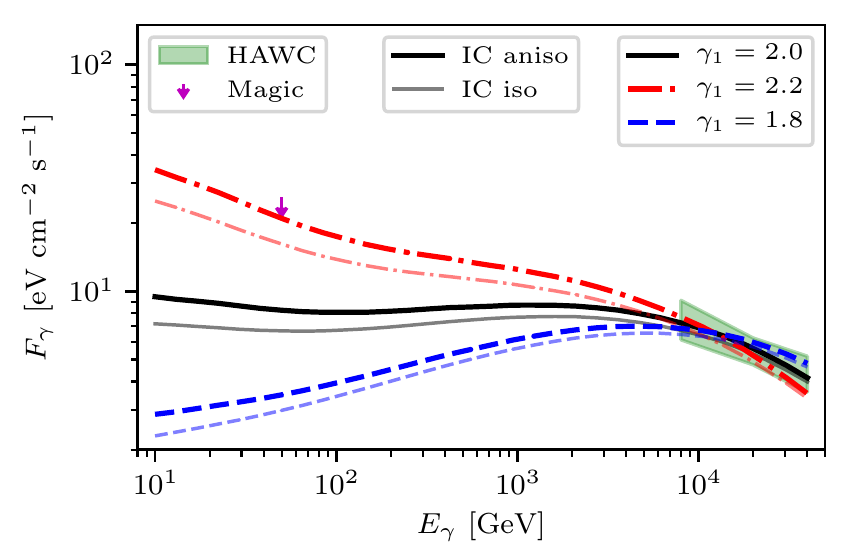}\\
  \includegraphics[width=0.48\textwidth]{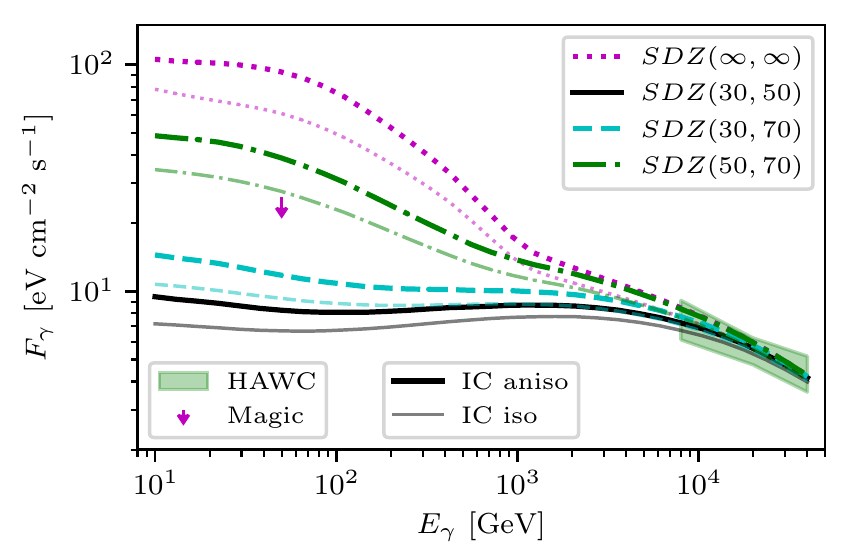}
  \caption{Spectrum of IC emission averaged over a $10^\circ$ wide region
    around the current location of Geminga. The models shown here assume
    Geminga is stationary and that the SDZ is stationary and centered at
    Geminga.  The top panel shows the spectrum for different values of
    $\gamma_1$ as indicated in the legend, but using fixed values of $(r_z,
    r_t)=(30,50)$~pc. The bottom panel shows the spectrum for models with
    different pairs of $(r_z, r_t)$, but fixed value of $\gamma_1=2.0$. The
    green shaded region corresponds to HAWC observations assuming a diffusion
    profile for the spatial distribution \citep{AbeysekaraEtAl:2017} and the
  magenta arrow is the upper limit from observations with the MAGIC telescope
  corrected for the diffusion profile \citep{TangPiran:2018}.}
  \label{fig:StationarySpectrum}
\end{figure}

Calculations with different values of $\gamma_1=1.8$ and 2.2 
are made using $(r_z, r_t)=(30,50)$~pc.  The conversion efficiency, $\eta$, is
correspondingly updated for these models to provide consistency with the HAWC data. The values are
$\eta = 0.18$ and $\eta = 0.75$ for $\gamma_1=1.8$ and $\gamma_1=2.2$,
respectively.  The average spectrum of the IC emission over a circular region
of $10^\circ$ in radius around the current location of Geminga is shown in
Figure~\ref{fig:StationarySpectrum}.  The models are all tuned to agree
with the HAWC data and, therefore, the predicted profiles differ significantly at lower energies.
Also shown is the prediction made using the isotropic assumption for the IC
cross section (dotted curve in upper panel), which is a commonly used approximation that ignores the angular dependence of the ISRF.
Using the isotropic approximation under-predicts the emission by $\sim$10\% at 10~TeV, and $\sim25$\% at 10~GeV.  The discrepancy between the IC
emission calculated with the realistic angular distribution of the ISRF and
with the isotropic distribution depends on the adopted electron injection spectrum
and the size of the SDZ, with softer injection spectra and larger SDZs producing larger differences. 

Kinematically, the energy dependence of the discrepancy can be understood where the very highest energy \gray{s} are produced in head-on collisions from back-scattered soft photons, where the ultrarelativistic electrons ``see'' the angular distribution of soft photons concentrated in a narrow (head on) beam. Meanwhile, for lower electron energies the angular distribution of the background photons is considerably more important and more significantly affects the energy of the upscattered \gray{s}.  
The softer injection spectra have more low-energy electrons and there is more confinement for low-energy electrons for larger sizes of the SDZ. 
The surface brightness profile is also slightly affected
by the isotropic assumption of the IC emission as shown in Figure~\ref{fig:StationarySB} (bottom panel).
The profile is more peaked
under the isotropic assumption; the under-prediction in the wings of the
profile is $\sim$ 10\% deeper than near the center.
Consequently, using the averaging made for the isotropic approximation has some effect that can produce an incorrect determination of the diffusion coefficient from the shape of the profile.

Also shown in
Figure~\ref{fig:StationarySpectrum} is the effect of the size of the SDZ on
the average spectrum.  A larger SDZ size results in a larger electron number density that, in turn, leads to an increased emission in the GeV energy range. The observations made with the MAGIC telescope
(\citealp{AhnenEtAl:2016}, updated to match the diffusion profile by
\citealp{TangPiran:2018}) constrain both the particle injection spectrum and
the size of the SDZ. A smaller SDZ and/or harder particle injection spectrum
is required by the observations.
The perceived degeneracy between the size of the SDZ and the
injected spectrum of the particles from this Figure is broken when the angular
profile of the low-energy emission is taken into account. Changing the injection spectrum
will not affect the shape of the angular profile while changing the SDZ size does. The angular extent of the emission should therefore
be a good indicator of the size of the SDZ.

\begin{figure}[tb]
  \centering
  \includegraphics[width=0.48\textwidth]{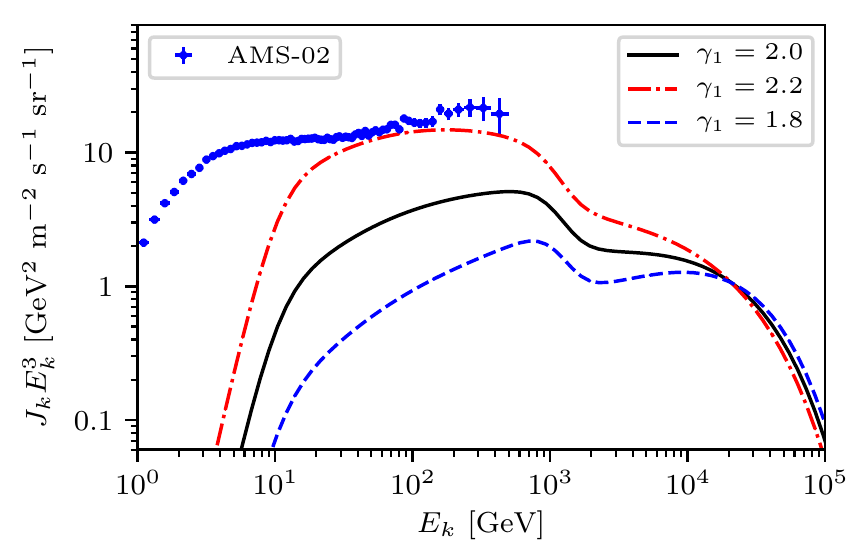}\\
  \includegraphics[width=0.48\textwidth]{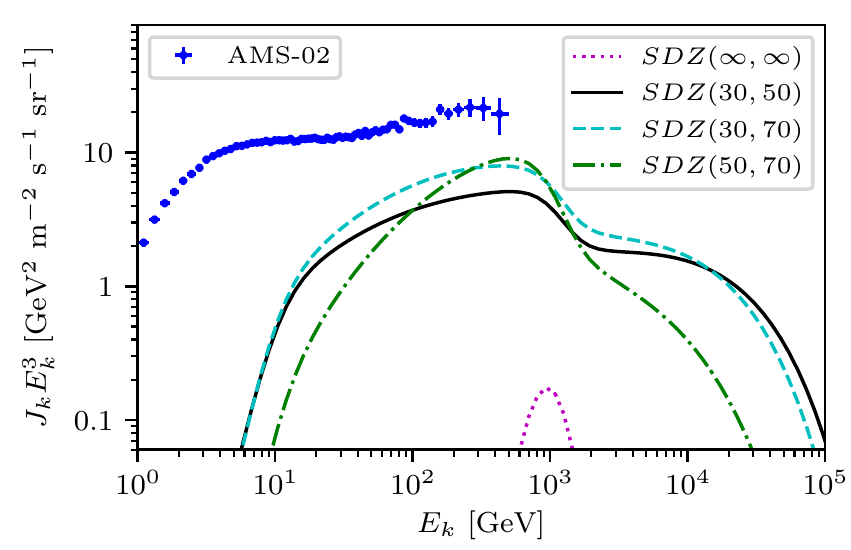}
  \caption{
Predicted flux of positrons at Earth. The models shown here assume Geminga is
stationary and that the SDZ is stationary and centered at the current location
of Geminga.  The top panel shows the spectrum for different values of
$\gamma_1$ as indicated in the legend, but using a fixed SDZ size $(r_z,
r_t)=(30,50)$~pc. The bottom panel shows the spectrum for models with different
pairs of $(r_z, r_t)$, but fixed value of $\gamma_1=2.0$. The points are
AMS-02 data \citep{AguilarEtAl:2014}.}
  \label{fig:StationaryPositronSpectrum}
\end{figure}

One of the main conclusions of the study by \citet{AbeysekaraEtAl:2017} is
that the unexpected rise in the positron fraction as measured by PAMELA
\citep{AdrianiEtAl:2009}, \fermilat{} \citep{AckermannEtAl:2012}, and AMS-02
\citep{2013PhRvL.110n1102A,AguilarEtAl:2014,2014PhRvL.113l1101A} cannot be due
to the positrons accelerated by the Geminga PWN. This conclusion is based on a
one-zone diffusion model constructed to agree with the HAWC data.  For a two-zone diffusion model the
conclusion can be quite different. \citet{ProfumoEtAl:2018} and
\citet{FangEtAl:2018} found that the positron excess could easily be
reproduced with such a model using a power-law injection spectrum for
positrons (and electrons) with index of $2.34$ and $2.2$, respectively.
\citet{TangPiran:2018} also used data from the MAGIC telescope to
constrain the model and found
that a harder particle injection spectrum below a break energy of 30~TeV was
necessary, which 
resulted in a calculated positron flux at Earth that did not fully explain the rise in the positron fraction.

Figure~\ref{fig:StationaryPositronSpectrum} shows the positron flux at Earth as predicted by the models considered in this paper,
where the top and bottom panels illustrate models with different SDZ sizes, and the effect of changing $\gamma_1$, respectively. The results clearly show a strong
variation in the expected positron flux at Earth.  A smaller SDZ
leads to larger flux of positrons at Earth because of the larger
\emph{effective} diffusion coefficient. Only if the SDZ extends all the way to
the Solar system are the results of \citet{AbeysekaraEtAl:2017} reproduced.  Larger
values of $\gamma_1$ also results in a larger positron flux at Earth in the observed
energy range, meanwhile values larger than $\gamma_1 \approx 2.2$ are excluded
as the predicted positron flux would exceed the data for a 
SDZ of size $(r_z,r_t)=(30,50)$~pc.

So far the calculations have been made considering a
stationary source located at the current position of Geminga $(l_G, b_G)$.
However, such an approximation is not supported by its observed large proper motion.
In the following analysis, the proper motion is taken into account, but different
assumptions are made on the origin of the SDZ and the value of
the drag coefficient $A_d$ introduced in Eq.~(\ref{eq:vDiffusion}).  The
different assumptions are referred to as scenarios A to D and detailed below.
For scenarios A and B, the SDZ is assumed to be stationary and is unrelated to
Geminga.  In both scenarios, $A_d=0$ and the pulsar velocity is constant.
Scenario A assumes that the SDZ is centered on the current position of
Geminga by chance,
and that the parameters of the zone are:
$(r_z,r_t)=(30,50)$~pc.  Scenario B assumes that the SDZ is centered at the
birth place of Geminga. In this scenario
the SDZ has to be much larger to extend from the birth place of Geminga to its
current location, $(r_z,r_t)=(90,110)$~pc. The
center of the spatial grid is placed at $(8.7358, 0.0962,
-0.0124)$~kpc for both these scenarios. This is about half-way between the birth place and the current
location of Geminga for the $x$- and $y$-axes, but only a quarter of the way
for the $z$-axis.  As described earlier, the location of the center of the grid is set to give the current location of Geminga preference over its birth location.

\begin{figure}[tb]
  \centering
  \includegraphics[width=0.48\textwidth]{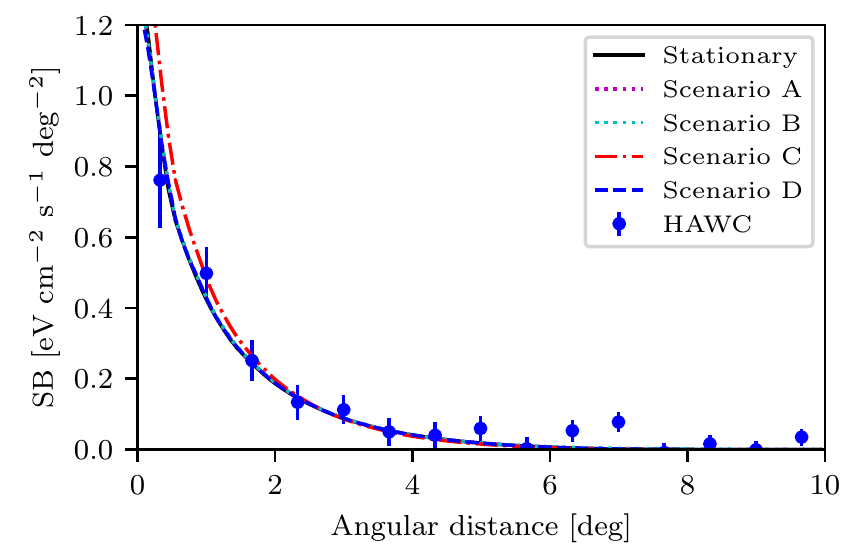}
  \caption{
Surface brightness of the modeled IC emission for scenarios A through D shown as a function of the angular distance from the current location of Geminga. Also shown is the profile for the model with a stationary Geminga PWN with $(r_z,r_t)=(30,50)$~pc and $\gamma_1=2.0$.  The data points show the profile observed by HAWC \citep{AbeysekaraEtAl:2017}.}
  \label{fig:ProfilesScenarios}
\end{figure}

\begin{figure*}[tb]
  \centering
  \includegraphics[width=0.49\textwidth]{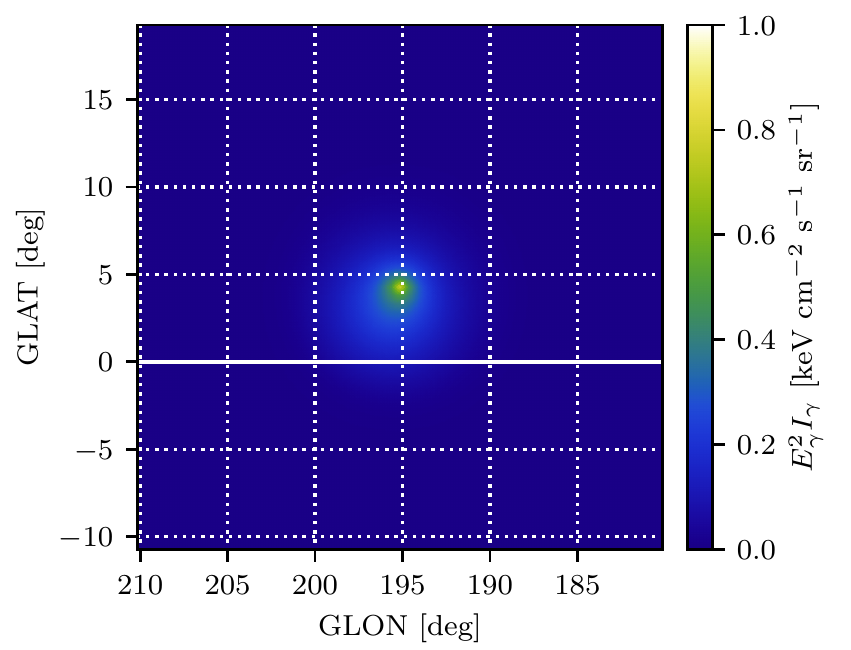}
  \includegraphics[width=0.49\textwidth]{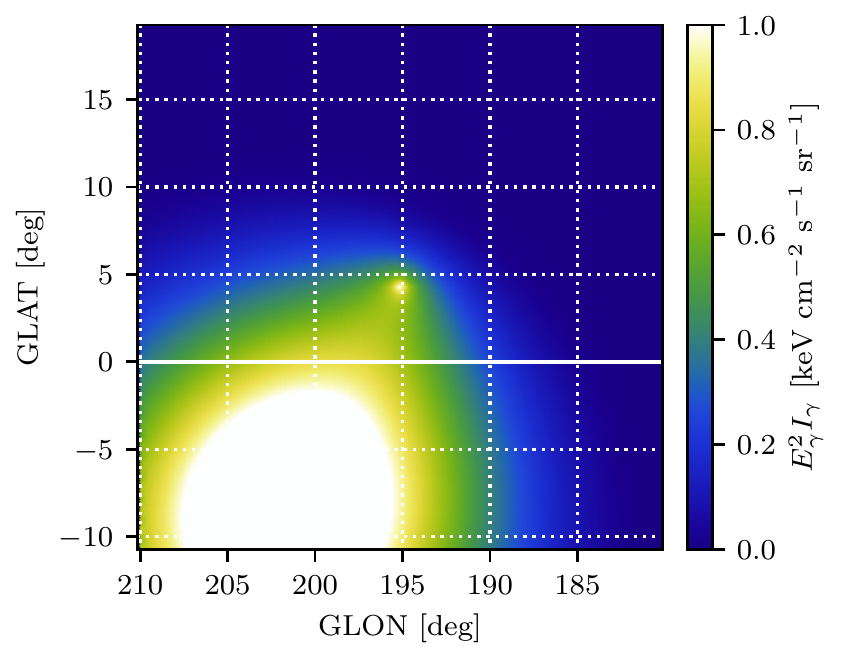}\\
  \includegraphics[width=0.49\textwidth]{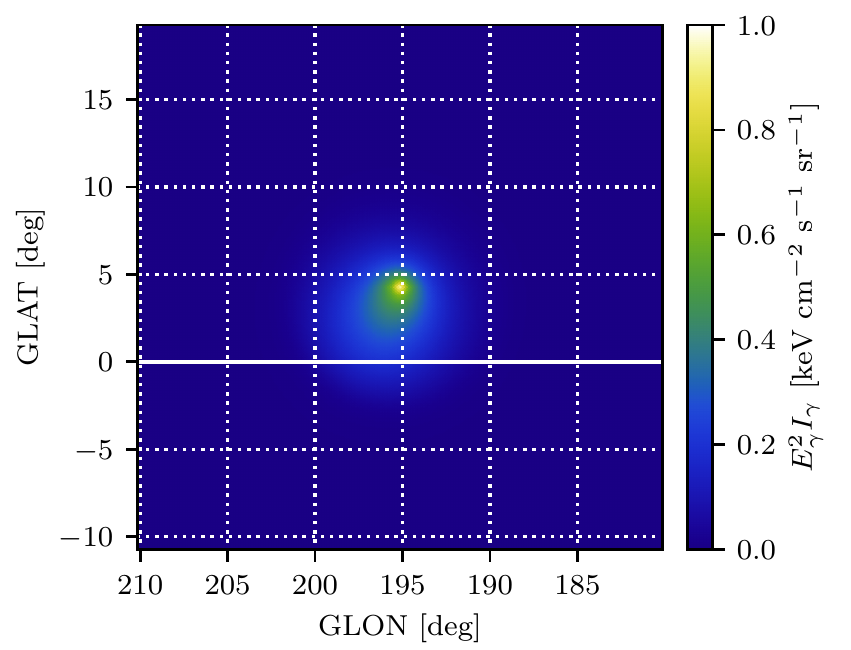}
  \includegraphics[width=0.49\textwidth]{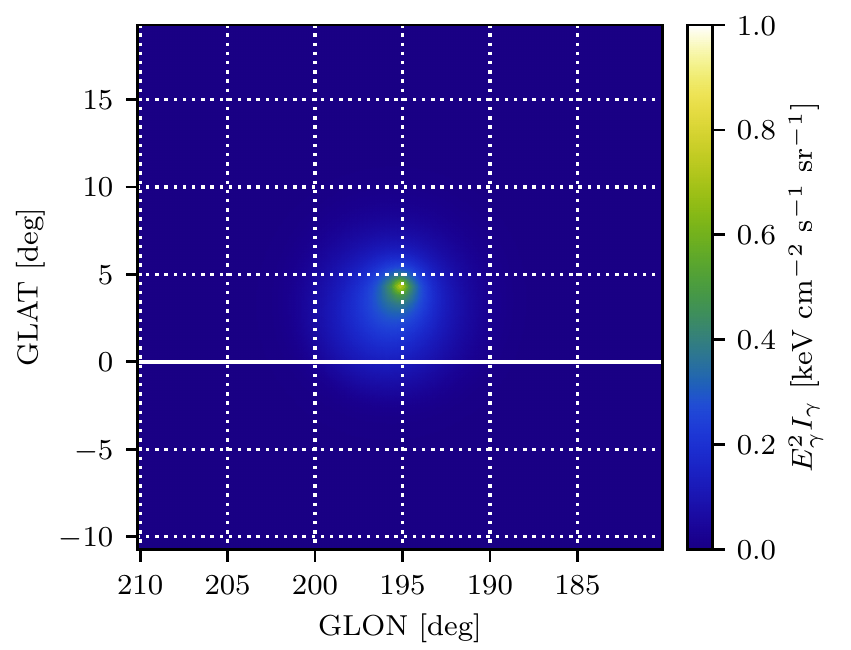}
  \caption{
IC intensity maps evaluated at 10~GeV around the current location of Geminga.  Shown are maps for scenarios A (top left), B (top right), C (bottom left), and D (bottom right).  The maps are displayed using the colormap bgyw from the colorcet package \citep{Kovesi:2015}.
  }
  \label{fig:Maps10GeV}
\end{figure*}

\begin{figure}[tb]
  \centering
  \includegraphics[width=0.48\textwidth]{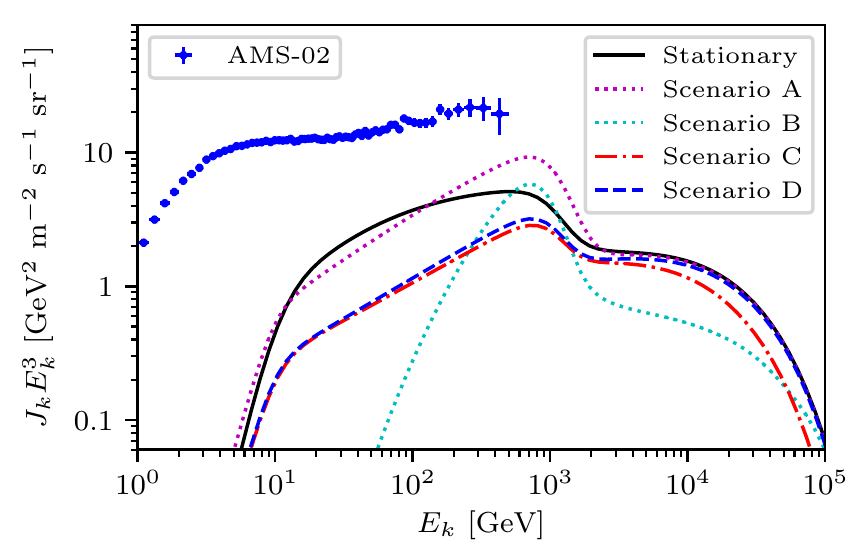}
  \caption{
Predicted positron fluxes at Earth for scenarios A through D. Also shown are the results for a model where Geminga is stationary with $(r_z,r_t)=(30,50)$~pc and $\gamma_1=2.0$.  The points are AMS-02 data \citep{AguilarEtAl:2014}.}
  \label{fig:PositronsScenarios}
\end{figure}

\begin{figure}[tb]
  \centering
  \includegraphics[width=0.48\textwidth]{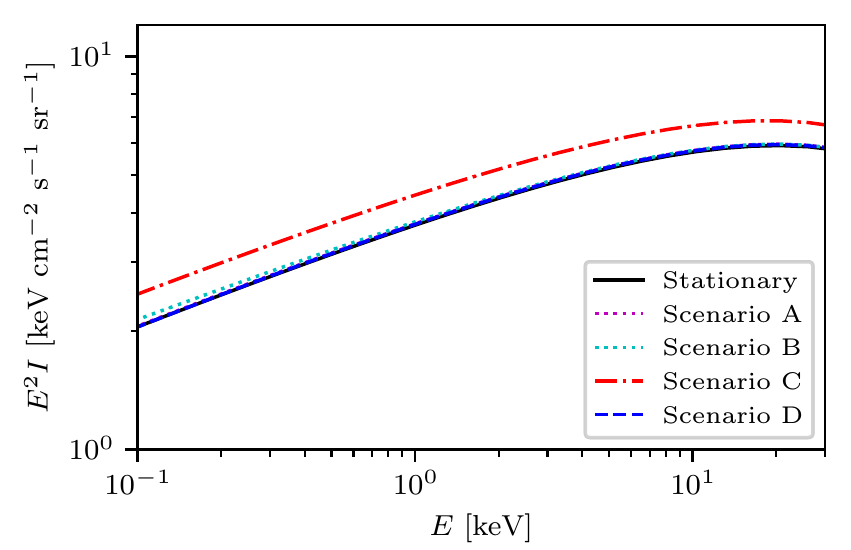}
  \caption{
Synchrotron intensity averaged over a region around the current location of Geminga with a radius of $1^\circ$ for scenarios A through D. Also shown are the
results for a model where Geminga is stationary with $(r_z,r_t)=(30,50)$~pc
and $\gamma_1=2.0$.}
  \label{fig:SynchrotronScenarios}
\end{figure}

For scenarios C and D, the center of the SDZ follows the location of Geminga
and its size increases proportionally to the square root of time, normalized such that the final size of the SDZ is
$(r_z,r_t)=(30,50)$~pc.  The difference between these two scenarios is the value
of the drag coefficient, $A_d$.  In scenario C, $A_d=0$ and the velocity is
constant over time, while in scenario D,
$A_d=2.636\times10^{-8}$~pc$^{-1}$~yr$^{-1}$ giving Geminga an initial
velocity of $410$~km~s$^{-1}$ at a
position of $(8.72775, 0.113017, -0.0787255)$~kpc.
For numerical stability the gradient of the diffusion coefficient is limited to be less than one order of magnitude per grid pixel.
Because the ratio $D_0/D_z \sim 300$, it is necessary to increase $D_z$ initially for scenarios C and D to fulfill this condition
until the difference between $r_t$ and $r_z$ corresponds to the size of 2 pixels in all directions of the grid.  Correspondingly, $D_z$
is increased by up to an order of magnitude for the first few thousand years
only, but this stability requirement does not significantly affect the results.
The center of the spatial grid in scenario C is the same as that in
scenarios A and B, while in scenario D the center of the grid is at $(8.7300,
0.1075, -0.0213)$~kpc to better resolve the entire evolution of Geminga.
Again the center is about half-way between the origin
and current location of Geminga for $x$- and $y$-axes, but only a quarter of
the way for the $z$-axis.  For scenarios A through D the variables $(r_t,
r_z)$ are fixed to the values provided above and $\gamma_1=2.0$. 

Figure~\ref{fig:ProfilesScenarios} shows the angular profile of the surface
brightness in the energy range 8~TeV to 40~TeV compared to the HAWC
observations \citep{AbeysekaraEtAl:2017}.  For comparison the results of the
model with stationary Geminga, $(r_z,r_t)=(30,50)$~pc and $\gamma_1=2.0$ are
also shown.  As expected, all models agree well with the data and
significantly overlap. Though it is not seen in the Figure, the emission is
still reasonably symmetric around the current location of Geminga.  To
  quantify the symmetry, the standard deviation of the distribution of
intensities of pixels in the calculated HEALPix maps within each angular bin
is calculated. The
standard deviation in each angular bin is of the order of 5\% to 10\% only.
Meanwhile, some differences between models are also noticeable: the largest
standard deviation of the emission core is seen in scenario D because of the
larger proper velocity of the PWN, while the largest deviation in the tail is
seen in scenario B with the largest SDZ size.

In contrast, at lower energies all models produce quite different
distributions of the surface brightness. In particular, considerable asymmetry in the intensity maps around 10 GeV is seen in
Figure~\ref{fig:Maps10GeV}, especially for scenario B with the largest SDZ.
All scenarios where Geminga is moving show a distinctive tail in the low
energy range.  In the extreme case of scenario B the tail is exceptionally
broad and the emission of the tail is, in fact, much brighter than that at the
current location of Geminga.  The differences between other scenarios are small
and hard to distinguish in these maps.  The tail in scenario C is slightly
more extended than in scenarios A and D.  Observations of the Geminga PWN at
10 GeV will thus hardly be able to distinguish between these scenarios,
but may be able to verify the presence of the tail in Scenario B.

Despite their similar \gray{} emission maps (with exception of
scenario B), all scenarios predict different positron fluxes at Earth
(Figure~\ref{fig:PositronsScenarios}). For energies below 1 TeV, the positron
flux in scenario A is about a factor of 5 larger than that predicted for
scenarios C and D.  The difference between scenarios C and D is much smaller
and mostly at the highest energies where scenario D produces a higher flux.
This is due to the faster movement of the diffusion zone in scenario D that
allows the CR particles to escape quickly once the diffusion zone has left
them behind. This is somewhat idealized and it is more likely that the SDZ
will have a shape elongated along the direction of motion. For such a case
the results would become closer to scenario B, with a longer IC emission tail
at low energies and smaller flux of positrons at low and high energies, but
with a larger peak at around 300~GeV.

In addition to IC emission, GALPROP can calculate the predicted synchrotron emission from the models.  In a magnetic field with a strength of several $\mu$G, as is expected in the vicinity of Geminga, the particles responsible for the TeV \gray-emission will also emit synchrotron photons with energies of $\sim$ keV.  Because of the shape of the electron/positron spectrum of the injected particles from Geminga, the emission in radio and mm wavelengths is significantly dimmer than that from CR electrons from the Milky Way at large.  At keV energies the synchrotron emission follows a profile similar to that of the TeV IC emission, being peaked at the current location of Geminga with a half-width of about a degree.  Figure~\ref{fig:SynchrotronScenarios} shows the spectrum of the average intensity of the synchrotron emission within a degree of the current location of Geminga.  All the models considered predict a similar intensity level of the emission that approximately follows a power-law $I\propto E^{-1.8}$.  The calculations are cut off at few 10s of keV because the electron spectrum can only be reliably determined up to the energies corresponding to the HAWC observations, anything beyond that is an extrapolation.  This energy is reached already at a few keV in the synchrotron spectrum.  The synchrotron emission is close to the level of the diffuse X-ray emission observed in the Milky Way ridge \citep{2000ApJ...534..277V}, but significantly less intense than that observed from the Geminga pulsar and the apparent nearby tails \citep{2010ApJ...715...66P}.

\section{Implications for Propagation Throughout the Milky Way}
\label{sec:MW}

If the result for the two PWNe reported in \citet{AbeysekaraEtAl:2017} are not special cases, then it is likely that there are similar pockets of slow diffusion around many CR sources elsewhere in the Milky Way. The effective diffusion coefficient would thus be smaller in regions where the number density of CR sources is higher. 
Secondary CR species with different inelastic production cross sections (e.g., $\bar{p}$ vs.\ B) probe different propagation distances.  Interpretations of their data would yield different average diffusion coefficients as was indeed found by \citet{JohannessonEtAl:2016}. 
Starburst galaxies with large star formation rate are expected to have very slow diffusion and should exhibit an energy-loss-dominated spectrum of \gray{} emission that is much flatter than that observed from galaxies where the leakage of CRs dominates the energy losses. Interestingly, exactly such evolution of the spectral shape is observed when galaxies with different star formation rate, such as the Magellanic Clouds, Milky Way, M31 vs.\ NGC 253, NGC 4945, M82, NGC 1068, are compared \citep[][and references therein]{2012ApJ...755..164A}.

Here it is assumed that the distribution of SDZs follows that of the CR sources, while the exact origin of these SDZs is not essential. With current CR propagation codes and reasonable resources resolving the entire Milky Way with a few pc grid size on short time scales is intractable.
A simpler approach must, therefore, be taken to explore the effects such SDZs have on the propagation of CRs throughout the Milky Way. Assuming that the properties of CR propagation and the distribution of CR sources vary on-average very little over the residence time of CRs in the Milky Way, a steady-state model can be used.  Also assuming that the residence time of CRs in the Milky Way is larger than the active injection time of the CR sources, the CR source distribution can be assumed smooth with injected power that approximates as constant with time.
These two approximations have been extensively used in the past and are the starting point of almost all studies on CR propagation across the Milky Way.

As was shown in the previous Section, the effect of an SDZ is a local increase of the density of CRs for a certain period of time, leading to an increase of CR reaction rates with the ISM contained within the SDZ. CRs thus spend more time in the vicinity of their sources compared to a model with a homogeneous diffusion.  This is equivalent to an effective decrease of the diffusion coefficient averaged over a larger volume. Even though, such an approximation does not account for the detailed spatial distribution of the individual SDZs, it does account for the increased interaction rate with the ISM, such as cooling and generation of secondary CR particles. Therefore, this approximation should still provide for the correct spectra and abundances of CR species. 

Assuming that CRs first propagate through a SDZ with diffusion coefficient $D_z$ and
radius $x_z$ that is embedded in a region with diffusion
coefficient $D_0$ and radius $x_0 \gg x_z$, their residence time in the total volume can be estimated as
\begin{equation}
  \tau \sim \frac{x_0^2}{D_0} + \frac{x_z^2}{D_z}.
  \label{eq:resTimeTwoZone}
\end{equation}
Assuming a single average diffusion coefficient $D$, the residence time can also be expressed as 
\begin{equation}
  \tau \sim \frac{x_0^2}{D}.
  \label{eq:resTimeAv}
\end{equation}
Combining the two equations results in
\begin{equation}
  D \approx D_0 \left[ 1 + \left( \frac{x_z}{x_0} \right)^{2} \frac{D_0}{D_z}\right]^{-1}.
  \label{eq:avDiffusion}
\end{equation}

Assuming that there is a SDZ around each CR source, the density of SDZs is
the same as the CR source number density, $q(\vec{x})$, where $\vec{x}$ is the spatial coordinate. Let $V_z$ be the volume of each SDZ and $\Delta V$ be the volume of an element in the spatial grid used in the calculation. Then for each volume element
\begin{equation}
  \frac{x_z}{x_0} \approx \left[\frac{V_z}{\Delta V} \int_{\Delta V} q(\vec{x}) dV\right]^{1/3} \approx \left[q_x V_z\right]^{1/3},
  \label{eq:xzx0approx}
\end{equation}
where $q_x$ is the CR source number density at the center of the volume element. The value $ Q=\int_V q(\vec{x}) dV$ is the total number of active CR sources in the Galaxy at any given time, and $QV_z$ is then the combined volume of all SDZs in the Galaxy. 
Assuming that the SN rate is 0.01 year$^{-1}$, and each source continuously accelerates particles for $\sim$$10^5$ years, there are $\sim$$10^3$ active CR sources at any given time. Assuming a radius of $x_z\sim 30$~pc for each source, the combined volume of all SDZs is $QV_z \approx 0.1$~kpc$^3$.
This value is in good agreement with estimates from \citet{HooperEtAl:2017} and
\citet{ProfumoEtAl:2018}.
The ratio $x_z/x_0$ in Eq.~(\ref{eq:xzx0approx}) depends on the product of the number of SDZs and their sizes.
Because of this, equivalent results can be obtained also for, e.g., fewer SDZs of larger sizes, provided that the total occupied volume is the same and their distribution follows $q(\vec{x})$.
Eq.~(\ref{eq:avDiffusion}) also depends on the value of $D_0/D_z$ which is set to 300 in the following  calculations, which is similar to the value used in the Geminga calculations. The results described below, therefore, are valid across different model assumptions provided they are consistent with Eq.~(\ref{eq:avDiffusion}).

\begin{figure*}[tb]
  \centering
  \includegraphics[width=0.49\textwidth]{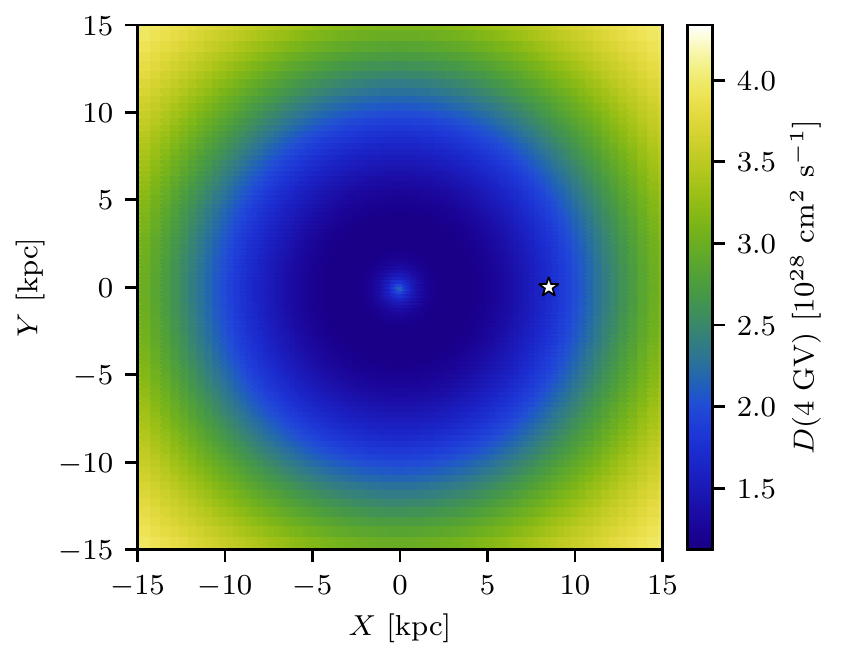}
  \includegraphics[width=0.49\textwidth]{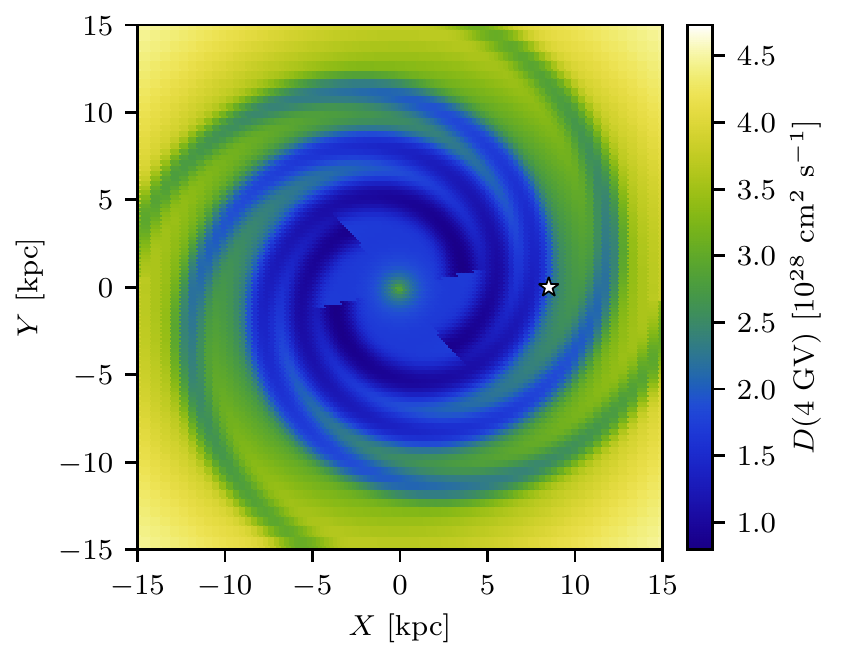}
  \caption{Diffusion coefficient evaluated in the Galactic plane at the normalization rigidity $R_0=4$~GV for a model where the effective diffusion coefficient is given by Eq.~(\ref{D_distr}). The left panel shows the distribution of the diffusion coefficient for the SA0 model while the right panel for the SA50 model.  The location of the Sun is marked as a white star and the GC is at (0,0). The maps are displayed using the colormap bgyw from the colorcet package \citep{Kovesi:2015}.
  }
  \label{fig:DiffusionPlane}
\end{figure*}

The modified diffusion described above has been implemented in the
GALPROP code to test its effect on the propagation of CR species in the Milky Way.
With a switch in the configuration file it is now possible to change to a diffusion configuration such that
\begin{equation}
    D(\vec{x}) = D_0 \left\{1 + \left[q\left( \vec{x} \right) V_z \right]^{2/3} \frac{D_0}{D_z} \right\}^{-1}.
    \label{D_distr}
\end{equation}
Therefore, even with $D_0$ and $D_z$ fixed the {\it effective} diffusion coefficient is still spatially varying, where the total volume of SDZs in the Milky Way is a normalization parameter $QV_z\approx 0.1$ kpc$^{3}$, as explained above. In principle, the number distribution of CR sources can be different from the distribution of their injected power into CRs in the Milky Way, but here they are assumed identical.
Two models SA0 and SA50 are used for the CR source density distribution
\citep{PorterEtAl:2017}.  The SA0 model is a 2D axisymmetric pulsar-like
distribution \citep{2004A&A...422..545Y} for the CR source density in the disk. The
SA50 model has half of the injected CR luminosity distributed as the SA0
density and the other half distributed as the spiral arms for the R12 ISRF
model.  Each CR source model is
paired with both a constant homogeneous diffusion and the modified diffusion
described above resulting in a total of four models.  The model calculations are performed on a 3D spatial grid that now includes the whole Milky Way using the grid function defined in Eq.~(\ref{eq:gridFunction}).  The parameters of the grid function are chosen such that the resolution in the $x$- and $y$-coordinates is 200~pc at the GC, increasing to 1 kpc at a distance of 20 kpc which is also the boundary of the grid.  At the distance of the Solar system the resolution is about 350~pc in the $x$-direction and the Sun is located at the center of a volume element.  In the $z$-direction the resolution is 50~pc in the plane, increasing to 200~pc at the boundary of the grid at $|z|=6$~kpc.  The energy grid is logarithmic, ranging from 3~MeV to 30~TeV with 112 energy planes.

The distribution of the effective diffusion coefficient $D(\vec{x})$ in the Galactic plane is shown in Figure~\ref{fig:DiffusionPlane}, where the numbers correspond to its value at the normalization rigidity $R_0=4$~GV.
The spiral arm structure of the SA50 model is clearly visible.
The maximum change in the effective diffusion coefficient is about a factor of 3, which corresponds to a peak value of $x_z/x_0 \sim 10^{-3}$.
All models employ the R12 ISRF from \citet{PorterEtAl:2017} and the 3D gas distributions from \citet{JohannessonEtAl:2018}.
The same standard procedure for parameter adjustment to match
the recent observations of CR species listed in Table~\ref{tab:CRdata} is followed for all source density models.
Solar modulation is treated using the force-field approximation \citep{1968ApJ...154.1011G}, one
modulation potential value for each observational period. The use of the latter is justified by the available resources and the main objective of this paper, i.e.\ to study the effect of the modified CR diffusion in a self-consistent manner, rather than to update the local interstellar spectra of CR species or to accurately determine the propagation parameters\footnote{For a detailed treatment of heliospheric propagation of CRs, see, e.g.,
  \citet{BoschiniEtAl:2017,BoschiniEtAl:2018b,BoschiniEtAl:2018a} and references therein.}.
The procedure for parameter tuning is the same as used by \citet{PorterEtAl:2017} and \citet{JohannessonEtAl:2018}.
The propagation parameters are first determined by fitting the models to the observed spectra of CR species, Be, B, C, O, Mg, Ne, and Si. These are then
kept fixed and the injection spectra for electrons, protons, and He fitted
separately. To reduce the number of parameters, only the most relevant primary abundances are fitted simultaneously with the propagation parameters while the rest are kept fixed. Because the injection spectra of all CR species are normalized relative to the proton spectrum at normalization energy 100 GeV/nucleon the procedure is repeated to ensure consistency. One iteration provides a satisfactory accuracy in all cases. The only difference from the procedure described in \citet{PorterEtAl:2017} and \citet{JohannessonEtAl:2018} is the inclusion of elemental spectra of Be, C, and O from AMS-02 and the elemental spectra from ACE/CRIS, while the data from HEAO-C3 and PAMELA has been removed.

\begin{deluxetable}{lll}[tb!]
\tablecolumns{3}
\tablewidth{0pc}
\tablecaption{Datasets used to derive propagation parameters
   \label{tab:CRdata}}
\tablehead{
\multicolumn{1}{l}{Instrument} &
\multicolumn{1}{l}{Datasets} &
\colhead{Refs.\tablenotemark{a}}
}
\startdata
AMS-02 (2011-2016) & Be, B, Be/B, Be/C, B/C, Be/O, B/O & I \\
AMS-02 (2011-2016) & C, O, C/O & II \\
AMS-02 (2011-2013) & e$^-$ & III \\
AMS-02 (2011-2013) & H & IV \\
AMS-02 (2011-2013) & He & V \\
ACE/CRIS (1997-1998) & B, C, O, Ne, Mg, Si & VI \\
Voyager 1 (2012-2015) & H, He, Be, B, C, O, Ne, Mg, Si & VII 
\enddata
\tablenotetext{a}{I: \citet{AguilarEtAl:2018}, II: \citet{AguilarEtAl:2017}, III:
\citet{AguilarEtAl:2014}, IV: \citet{AguilarEtAl:2015a}, V:
\citet{AguilarEtAl:2015b}, VI: \citet{GeorgeEtAl:2009}, VII:
\citet{CummingsEtAl:2016}}
\end{deluxetable}

\begin{deluxetable}{lcccc}[tb!]
\tablecolumns{5}
\tablewidth{0pc}
\tablecaption{Final propagation model parameters. \label{tab:CRparameters} }
\tablehead{
   & \multicolumn{2}{l}{Homogeneous diffusion} & \multicolumn{2}{l}{Modified diffusion} \\
\multicolumn{1}{l}{Parameter} & 
\colhead{SA0} & 
\colhead{SA50} &
\colhead{SA0} & 
\colhead{SA50} 
}
\startdata
\tablenotemark{a}$D_{0,xx}$ [$10^{28}$cm$^2$\,s$^{-1}$] & $4.36$            & $4.55$            
                                                       & $4.41$            & $4.78$            \\
\tablenotemark{a}$\delta$                              & $0.354$           & $0.344$           
                                                       & $0.358$           & $0.360$           \\
$v_{A}$ [km s$^{-1}$]                                  & $17.8$            & $18.1$            
                                                       & $15.7$            & $17.6$            \smallskip\\
\tablenotemark{b}$\gamma_0$                            & $1.33$            & $1.43$            
                                                       & $1.40$            & $1.47$            \\
\tablenotemark{b}$\gamma_1$                            & $2.377$           & $2.399$           
                                                       & $2.403$           & $2.381$           \\
\tablenotemark{b}$R_1$ [GV]                            & $3.16$            & $3.44$            
                                                       & $3.80$            & $3.82$            \\
\tablenotemark{b}$\gamma_{0,p}$                        & $1.96$            & $1.99$            
                                                       & $1.92$            & $1.93$            \\
\tablenotemark{b}$\gamma_{1,p}$                        & $2.450$           & $2.466$           
                                                       & $2.469$           & $2.453$           \\
\tablenotemark{b}$\gamma_{2,p}$                        & $2.391$           & $2.355$           
                                                       & $2.359$           & $2.321$           \\
\tablenotemark{b}$R_{1,p}$ [GV]                        & $12.0$            & $12.2$            
                                                       & $11.3$            & $11.3$            \\
\tablenotemark{b}$R_{2,p}$ [GV]                        & $202$             & $266$             
                                                       & $213$             & $371$             \\
$\Delta_{\rm He}$                                          & $0.033$           & $0.035$           
                                                       & $0.039$           & $0.032$           \smallskip\\
\tablenotemark{b} $\gamma_{0,e}$                       & $1.62$            & $1.49$            
                                                       & $1.46$            & $1.46$            \\
\tablenotemark{b} $\gamma_{1,e}$                        & $2.843$           & $2.766$           
                                                       & $2.787$           & $2.762$           \\
\tablenotemark{b} $\gamma_{2,e}$                       & $2.494$           & $2.470$           
                                                       & $2.506$           & $2.480$           \\
\tablenotemark{b} $R_{1,e}$ [GV]                        & $6.72$            & $5.14$            
                                                       & $5.03$            & $5.13$            \\
\tablenotemark{b} $R_{2,e}$ [GV]                        & $52$              & $69$            
                                                       & $69$              & $69$              \smallskip\\
\tablenotemark{c}$J_p$  & $4.096$ & $4.113$ 
                                                       & $4.102$           & $4.099$           \\
\tablenotemark{c}$J_e$  & $2.386$ & $2.362$ 
                                                       & $2.288$           & $2.345$           \smallskip\\
\tablenotemark{d}$q_{0,^{4}\rm He} \hfill [10^{-6}]\qquad\qquad$          & $ 92495$          & $ 91918$          
                                         & $ 92094$          & $ 93452$          \\
\tablenotemark{d}$q_{0,^{12}\rm C} \hfill [10^{-6}]\qquad\qquad$           & $  2978$          & $  2915$          
                                         & $  2912$          & $  2986$          \\
\tablenotemark{d}$q_{0,^{16}\rm O} \hfill [10^{-6}]\qquad\qquad$           & $  3951$          & $  3852$          
                                                               & $  3842$          & $  3956$          \\
\tablenotemark{d}$q_{0,^{20}\rm Ne} \hfill [10^{-6}]\qquad\qquad$          & $   358$          & $   327$          
                                          & $   322$          & $   359$          \\
\tablenotemark{d}$q_{0,^{24}\rm Mg} \hfill [10^{-6}]\qquad\qquad$          & $   690$          & $   704$          
                                          & $   681$          & $   744$          \\
\tablenotemark{d}$q_{0,^{28}\rm Si} \hfill [10^{-6}]\qquad\qquad$          & $   801$          & $   786$          
                                                               & $   762$          & $   833$          \smallskip\\
\tablenotemark{e}$\Phi_{\rm AMS,I}$ [MV]                  & $ 729$            & $ 741$            
                                                       & $ 709$            & $ 729$            \\
\tablenotemark{e}$\Phi_{\rm AMS,II}$ [MV]                 & $ 709$            & $ 729$            
                                                       & $ 696$            & $ 729$            \\
\tablenotemark{e}$\Phi_{\rm ACE/CRIS}$ [MV]               & $ 359$            & $ 370$            
                                                       & $ 345$            & $ 354$           
\enddata
\tablenotetext{a}{$D(R) \propto \beta R^{\delta}$, $D_0$ is the normalization at $R_0=4$~GV. 
}
\tablenotetext{b}{The injection spectrum is parameterized as $q(R) \propto R^{-\gamma_0}$ for $R \le R_1$, $q(R) \propto R^{-\gamma_1}$ for $R_1 < R \le R_2$, and $q(r) \propto R^{-\gamma_2}$ for $R > R_2$.  The spectral shape of the injection spectrum is
the same for all species except CR $p$ and He. $R_1$, and $R_2$ are the same for $p$
and He and $\gamma_{i,\rm He} = \gamma_{i,p}-\Delta_{\rm He}$}
\tablenotetext{c}{The CR $p$ and e$^-$ fluxes are normalized at the Solar
location at a kinetic energy of 100~GeV.  $J_p$ is in units of $10^{-9}$
cm$^{-2}$ s$^{-1}$ sr$^{-1}$ MeV$^{-1}$ and $J_e$ is in units of $10^{-11}$ cm$^{-2}$ s$^{-1}$ sr$^{-1}$ MeV$^{-1}$.}
\tablenotetext{d}{The injection spectra for isotopes are normalized relative to the proton injection spectrum at 100~GeV/nuc.  
The normalization constants for isotopes not listed here are the same as given in \citet{JohannessonEtAl:2016}.}
\tablenotetext{e}{The force-field approximation is used for calculations of the solar
modulation and is determined independently for each model and each
observing period. $\Phi_{\rm AMS,I}$ and $\Phi_{\rm AMS,II}$ correspond
to the 2011-2016 and 2011-2013 observing periods for the AMS-02
instrument, respectively.}
\end{deluxetable}

Table~\ref{tab:CRparameters} lists the best-fit parameters for the four models
considered.  The parameters are similar for all models and the modified
diffusion and different source distributions only slightly affect the best-fit
values.
Most interestingly, despite the fact that the value of the effective diffusion coefficient exhibits strong spatial variations in the Galactic plane including at the Solar system location, where it is about a factor of 2 smaller compared to that in the model with homogeneous diffusion (Fig.~\ref{fig:DiffusionPlane}), the value of $D_0$ is not significantly different from that obtained for the homogeneous model.
This is connected with the relatively small volume affected by the modified diffusion regions because they are associated with the CR sources, which have a relatively narrow distribution about the Galactic plane with exponential $z$ scale-height 200~pc.
The modified diffusion slightly affects the low-energy part of the spectrum, resulting in smaller
modulation potentials and Alfv{\'e}n speeds.
Overall the addition of the SDZs does not significantly alter the global diffusive properties of the ISM.

\begin{figure*}[tb]
  \centering
  \includegraphics[width=0.49\textwidth]{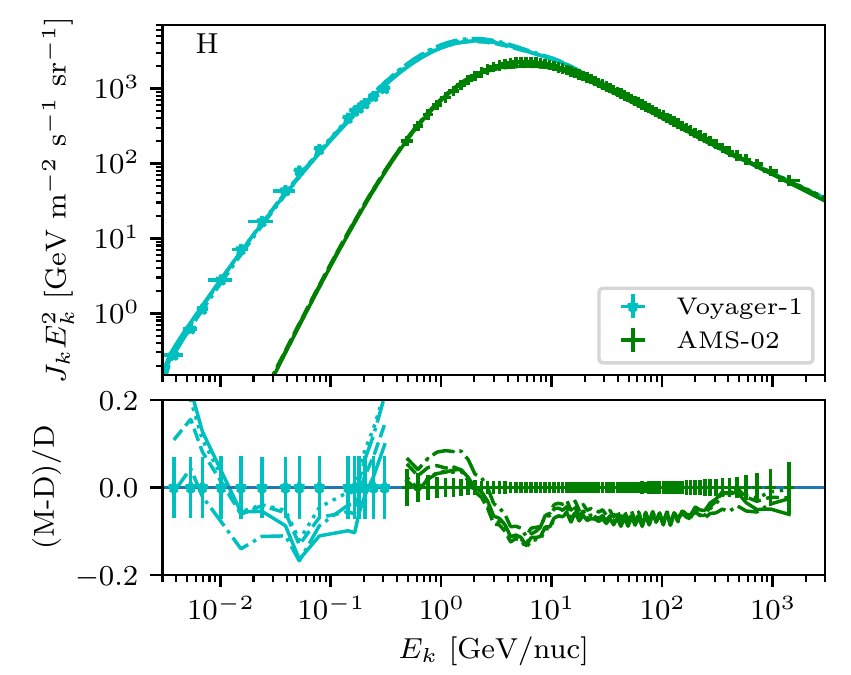}
  \includegraphics[width=0.49\textwidth]{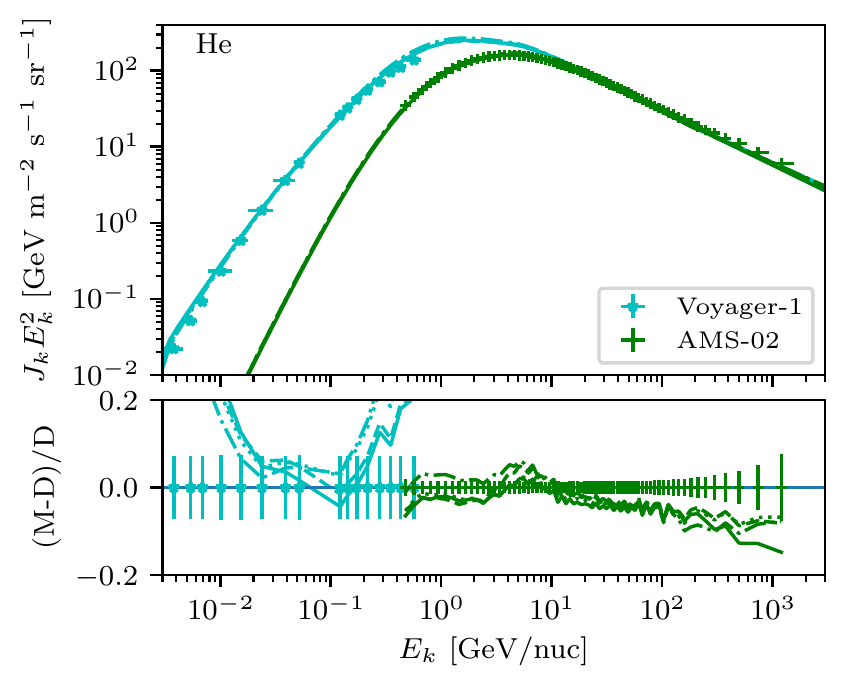}\\
  \includegraphics[width=0.49\textwidth]{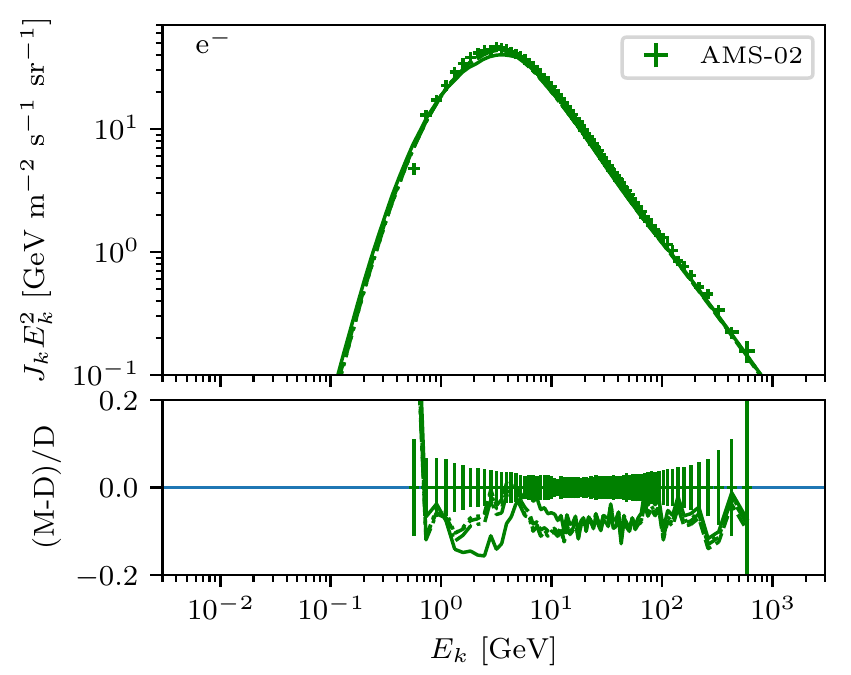}
  \includegraphics[width=0.49\textwidth]{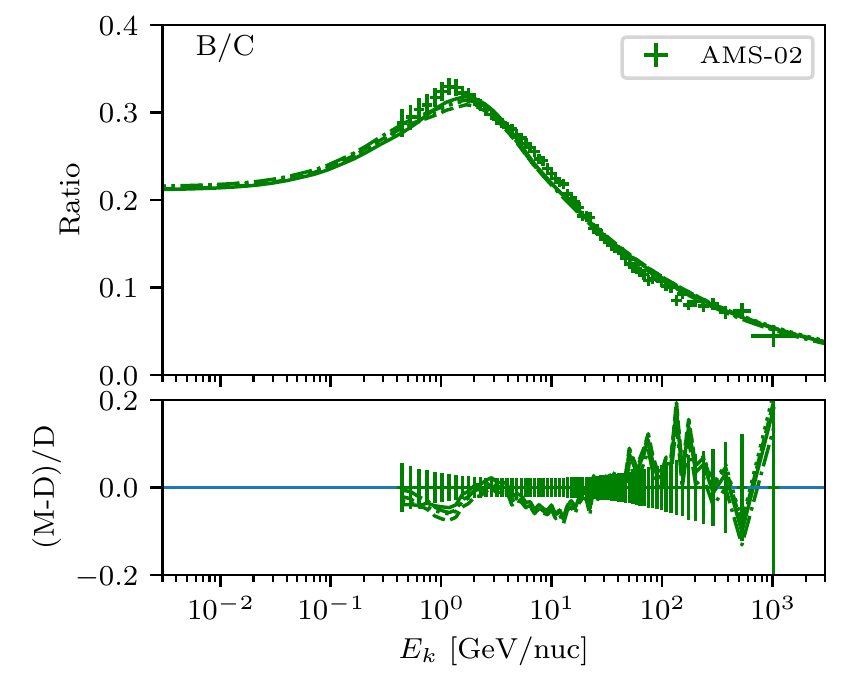}\\
  \includegraphics[width=0.49\textwidth]{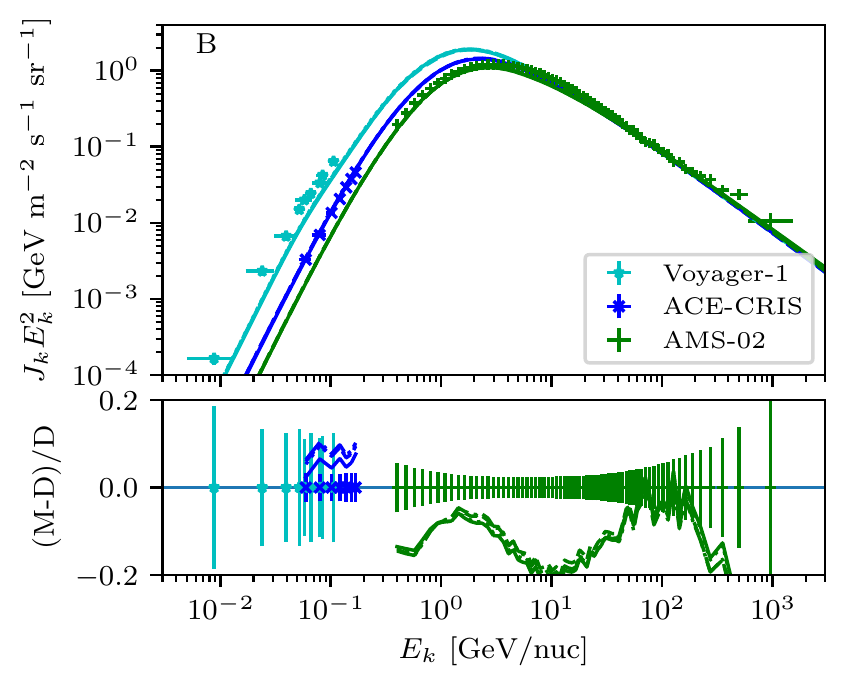}
  \includegraphics[width=0.49\textwidth]{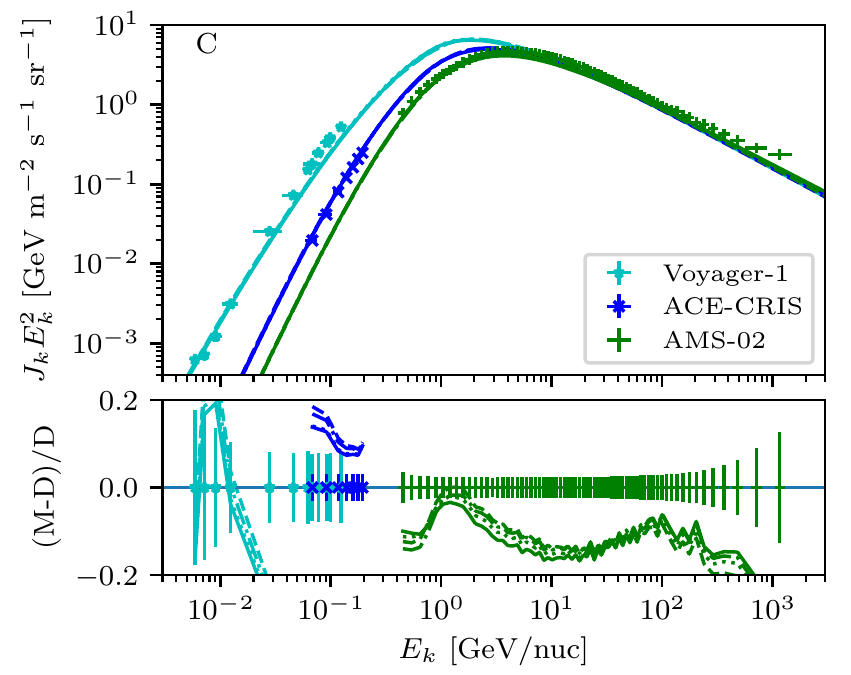}\\
  \caption{A comparison between the best-fit model predictions and observation of CR species
near the Solar system. The bottom panel of each subfigure shows the data subtracted from the model in units of data flux. References to the data are provided in Table~\ref{tab:CRdata}. For uniformity of the presentation, AMS-02 data has been converted from rigidity to kinetic energy per nucleon units using isotopic composition of each element as measured by ACE/CRIS, but the likelihood in the fitting procedure is calculated using rigidity. Different curves represent the four different models considered.  SA0 and SA50 cases with the homogeneous diffusion are shown as solid and dotted lines, respectively, while SA0 and SA50 cases with the modified diffusion are show as dashed and dash-dotted lines, respectively. Note that there is significant overlap between the lines for the different models for the top panels of the subfigures.
  }
  \label{fig:CRelements}
\end{figure*}

A comparison between the calculated spectra of CR species and data is shown in Figure~\ref{fig:CRelements}.  The
models agree reasonably well with the data and deviate less than 10\% from the
measurements for most elements.  The largest systematic deviations
shown are at around 10~GeV for B and C where the relative residuals are as large as
$\sim$20\%.  The residuals for Be and O are similar (not shown).  These
are likely caused by the lack of freedom in the injection spectra for the
primaries, because the deviations are hardly visible in the secondary-to-primary 
ratios of which B/C is shown in the Figure for illustration.  If the deviations were caused by propagation
effects they would be stronger in the secondary component and be
correspondingly more
prominent in the secondary-to-primary ratios.  Deviations from the H
and He data are smaller and the feature at 10~GeV is not visible.
Figure~\ref{fig:CRelements} also shows that the models agree with each other to
within 5\% for all elements, and this further illustrates that the addition of the SDZ pockets does not strongly affect predictions for the local spectra of CR species.

\section{Discussion}

Observations of \gray{} emission from two PWNe made by HAWC \citep{AbeysekaraEtAl:2017} provide a direct insight into the diffusive properties of the ISM on small scales. Combined with direct measurements of the spectra of CR species in the Solar neighborhood, the HAWC observations point to quite different diffusion coefficients in space surrounding the Geminga and PSR~B0656+14 PWNe compared to the ISM at large, with that associated with the former (SDZs) about two orders of magnitude smaller than the latter. The size of the SDZs is constrained by the upper limits of \gray{} emission at lower energies obtained by the MAGIC telescope to be $\lesssim 100$~pc \citep{AhnenEtAl:2016, TangPiran:2018}.  Observations by the \fermilat{} may be used to further constrain the properties of the SDZs, although it may be difficult to break a degeneracy between the size of the diffusion zone and the shape of the injection spectra of CR electrons and positrons.  This is particularly true for PWNe with a large proper motion, like the case of Geminga, which significantly affects the shape of the IC emission in the energy range of the \fermilat.

One of the main claims of \citet{AbeysekaraEtAl:2017} is that positrons produced by Geminga provide a negligible fraction of the positrons observed by the AMS-02 instrument \citep{AguilarEtAl:2014}.  This statement is correct only if the diffusion coefficient is small in the entire region between Geminga and the Solar system.  Using a two-zone model consistent with both HAWC and MAGIC observations of Geminga, the predicted positron flux can vary over a wide range from a small fraction to almost the entire observed positron flux, dependent on the assumed properties of the SDZ and the injection spectrum.  Even if the \gray{} emission can be constrained with the \fermilat{} observations, changing the SDZ properties can result in a variation of the predicted flux by a factor of few.  This is illustrated by Scenarios A and C shown in Figures~\ref{fig:Maps10GeV} and~\ref{fig:PositronsScenarios}: while the predicted \gray{} emission differs only marginally, the predicted positron flux differs by about a factor of 5.

If the result for the two PWNe reported in \citet{AbeysekaraEtAl:2017} are not special cases, then it is likely that there are similar pockets of slow diffusion around many CR sources elsewhere in the Milky Way. In the likely case that the distribution of the SDZs follows the distribution of CR sources, CRs spend more time in the inner Milky Way and generally in the plane than in the outer Milky Way or its halo.
Despite this the predicted fluxes of secondary CR species near the Solar system are not strongly affected (see Figure~\ref{fig:CRelements}), and the propagation model parameters obtained with SDZs included are very close to those determined for models using homogeneous diffusion (see Table~\ref{tab:CRparameters}). This is a non-trivial result because the production rate of secondary CR species may also increase in the same regions of slow diffusion provided there is enough interstellar gas there. Compared to a model with homogeneous diffusion, the density of CRs should increase towards the inner Milky Way where the distribution of CR sources peaks. Correspondingly, the interstellar high-energy \gray{} emission (IEM) should also be brighter in the direction of the inner Milky Way. At the same time, this differs from the results of the analysis of the \fermilat{} data which indicate that the gradient is even smaller than predicted by models with two-dimensional axisymmetric geometry and homogeneous diffusion \citep[e.g.][]{AbdoEtAl:2010,AckermannEtAl:2011}. An updated analysis that uses three-dimensional spatial models and includes localized propagation effects may lead to a new interpretation of the distribution of the diffuse emission and of the \fermilat{} data.

In this study the GALPROP framework has been used, which is fully capable of calculating the spectrum and distribution of the expected interstellar diffuse \gray{} emission, but because of the assumptions incorporated into the modified diffusion model, the predictions of the diffuse emission would be impractical. The sources of CRs are transient in nature and spatially localized and so are the potential SDZs associated with them. Depending on the exact nature of the SDZs, CR particles  may be confined within these regions for a significant fraction of their residence time in the Milky Way, contrary to the assumption made in Eqs.~(\ref{eq:avDiffusion}) and (\ref{eq:xzx0approx}). This may lead to the incorrect brightness distribution of the predicted \gray{} emission, which is sensitive to the details of CR distribution throughout the Milky Way. Properly accounting for the transient nature of CR sources in the entire Milky Way is beyond the scope of this work, but will be addressed  by a forthcoming paper.

One plausible explanation for the generation of SDZs near CR sources is the self-excitation of Alfv{\'e}n waves by the CRs themselves as they stream out into the ISM. Such streaming instabilities have been discussed since the early 70s \citep{Skilling:1971} and are reviewed in \citet{AmatoBlasi:2018}. The size and magnitude of the effect of streaming depends on the gradient of the CR distribution and the properties of the ISM, in particular the number density of neutrals and the Alfv{\'e}n speed. Numerical studies of CR diffusion near SNRs reveal that lower number densities lead to smaller sizes for the SDZ, but also smaller change in the diffusion constant \citep{NavaEtAl:2016,NavaEtAl:2019}. It is also expected that the injection spectrum is time dependent and may lead to an energy dependence of the diffusion coefficient in SDZs that is quite different from that in the ISM. Such models for the SDZ around Geminga would be in agreement with the HAWC observation that only constrains the diffusion at the highest energies, but produce a wide range of predictions for the expected flux of CR positrons near the Solar system as well as for the expected spectrum of IC emission at low energies. In turn, this would result in an increased CR flux in the inner Galaxy that is dominated by the high-energy particles, leading to a hardening of the spectrum that has been indeed observed by the \fermilat{} \citep{SeligEtAl:2015,AjelloEtAl:2016}. Exploration of these effects is deferred to future work.

To summarize, observations of the SDZs made by HAWC provide an interesting opportunity to get insight into the fairly complex details of the CR propagation phenomenon. Extension of these observation onto the whole Milky Way \emph{in specialibus generalia quaerimus} is, however, a non-trivial task 
and further understanding of the origin and properties of the SDZs are necessary to get a correct picture.

\acknowledgements
T.A.P. and I.V.M. acknowledge partial support from NASA grant NNX17AB48G.

\bibliography{biblio}

\end{document}